\documentclass{article}

\usepackage[
top=3cm,
bottom=3cm,
headheight=40pt,
headsep=20pt,
a4paper]{geometry}

\usepackage{fancyhdr}       
\usepackage{XCharter}       
\usepackage[scaled=1.1]{zlmtt} 
\usepackage[colorlinks, breaklinks=true, citecolor=black]{hyperref}
\usepackage{graphicx}
\usepackage{natbib}
\usepackage{xspace}

\fancypagestyle{firstpage}{
  \fancyhf{} 
  \fancyhead[L]{\includegraphics[width=70pt]{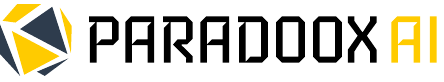}}
  \fancyhead[R]{FinLLM Workshop at IJCAI 2025}
  \fancyfoot[C]{\thepage}

}

\newcommand{\sys}{\texttt{\textbf{DeepFund}}\xspace}

\title{Will LLMs be Professional at Fund Investment? \\ DeepFund: A Live Arena Perspective}

\author{%
  Changlun Li$^{1,2}$~
  Yao Shi$^{1,2}$~
  Yuyu Luo$^{1}$~
  Nan Tang$^{1}$\thanks{
  Corresponding author: nantang@hkust-gz.edu.cn
  } \\
  $^1$The Hong Kong University of Science and Technology (Guangzhou) \\
  $^2$PAI Lab
}

\date{}

\begin{document}

\maketitle
\thispagestyle{firstpage}

\begin{abstract}
Large Language Models (LLMs) have demonstrated impressive capabilities across various domains, but their effectiveness in financial decision-making remains inadequately evaluated. 
Current benchmarks primarily assess LLMs' understanding on financial documents rather than the ability to manage assets or dig out trading opportunities in dynamic market conditions. 
Despite the release of new benchmarks for evaluating diversified tasks on the financial domain, we identified four major problems in these benchmarks, which are data leakage, navel-gazing, over-intervention, and maintenance-hard. 
To pave the research gap, we introduce DeepFund, a comprehensive arena platform for evaluating LLM-based trading strategies in a live environment. Our approach implements a multi-agent framework where they serve as multiple key roles that realize the real-world investment decision processes. 
Moreover, we provide a web interface that visualizes LLMs' performance with fund investment metrics across different market conditions, enabling detailed comparative analysis. 
Through DeepFund, we aim to provide a more realistic and fair assessment on LLM's capabilities in fund investment, offering diversified insights and revealing their potential applications in real-world financial markets. Our code is publicly available at \url{https://github.com/HKUSTDial/DeepFund}.
\end{abstract}

\section{Introduction}

The financial industry has witnessed a revolution with the integration of artificial intelligence technologies over the past decade~\cite{li2023large,Liu2024ASO,zhao2024revolutionizing}. From high-frequency trading algorithms~\cite{Liu2020AdaptiveQT,Briola2021DeepRL,Zong2024MacroHFTMA,liu2020finrl} to risk assessment models~\cite{giudici2018fintech,zheng2019finbrain,javaid2024ai}, AI has transformed how financial institutions operate and make decisions. Recently, Large Language Models (LLMs) have emerged as particularly promising tools for financial analysis because of their ability to process vast amounts of contextual information, understand market sentiment, and generate insights from diverse data sources~\cite{yang2023fingpt,zhang2023instructfingpt,yu2023temporal,wu2023bloomberggpt,yang2024finrobot}.
Furthermore, the Agentic AI technique is increasingly being deployed in autonomous trading flow, which can make investment decisions based on multi-modal market information, such as regional policy, earnings reports, technical indicators, etc~\cite{xiao2024tradingagents,wang2024quantagent,yu2024finmem,zhang2024multimodal,li2025hedgeagents}.

For the sake of a comprehensive evaluation on LLM-based approaches in the field of Financial AI, researchers have conducted a series of benchmarking studies.
Current evaluation benchmarks such as TAT-QA~\cite{Zhu2021TATQAAQ}, FinanceBench~\cite{islam2023financebench}, CFBenchmark~\cite{lei2023cfbenchmark} and FinNLI~\cite{Magomere2025FinNLIND} have made valuable contributions by assessing LLMs' understanding of financial documents, terminology, and basic reasoning. For example, LLM is required to answer questions about financial documents, such as balance sheets, income statements or financial news articles. 
However, these benchmarks primarily focus on the representation learning of financial knowledge rather than exploring models' abilities to make effective investment decisions in dynamic market conditions. 
Therefore, several modern benchmarks gradually shift their focus to the decision-making tasks, such as FinRL-Meta~\cite{Liu2022FinRLMetaME}, FinBen~\cite{xie2024finben} and InvestorBench~\cite{li2024investorbenchab}.

\begin{figure}[t!]
    \centering
    \includegraphics[width=.9\textwidth]{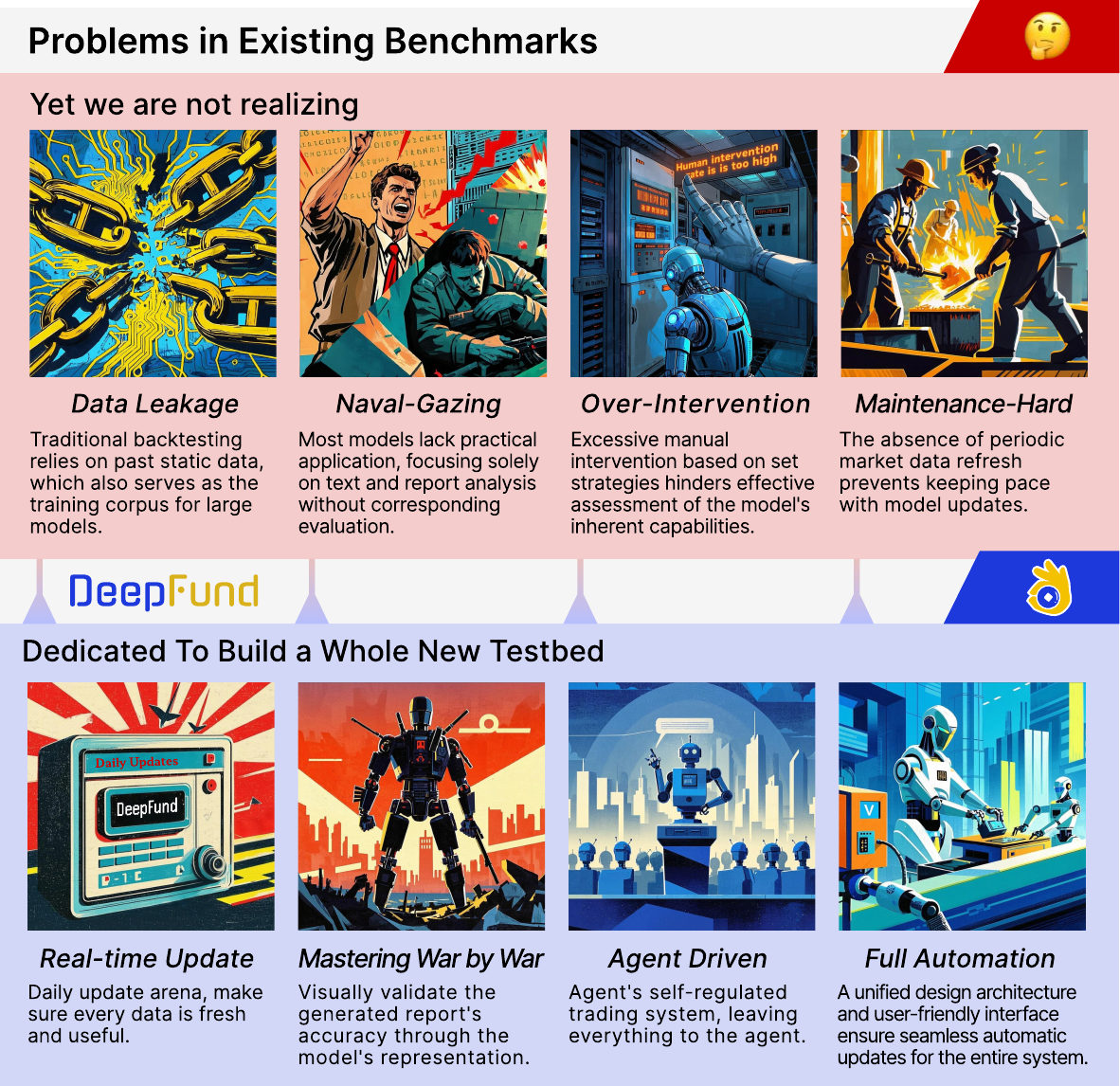}
    \caption{The problem in existing solutions and our vision.}
    \label{fig:problem}
\end{figure}

A primary limitation in existing benchmarks lies in the backtest methodology—the standard method for assessing trading strategies. This approach is particularly problematic when applied to LLMs, as these models have likely been pre-trained on substantial portions of financial history~\cite{petroni2019language,Liu2021PretrainPA}. As revealed in technical reports, different LLMs have varying knowledge cutoffs—DeepSeek-R1's training data ends in December 2023~\cite{guo2025deepseek}, while GPT-4o's extends to June 2024~\cite{hurst2024gpt}. When an LLM trained on data up to a certain date is evaluated on historical market data from before that date, \textbf{information leakage}~\cite{Xu2024BenchmarkingBL,ravaut2024much,ni2025training} becomes inevitable, creating misleading assessments of the model's true predictive capabilities.
Apart from that, we discuss more problems in Section~\ref{sec:problem}.

To address these limitations, we introduce \sys, a comprehensive arena platform for evaluating LLM-based trading strategies in a live environment. Our approach differs from previous solutions in several contributions:

\begin{itemize}
    \item We construct a holistic arena platform that mimics real-time trading environment, mitigating the information leakage problem inherent in traditional backtesting.
    \item We implement a multi-agent framework where LLMs serve as trading planner, analysts and managers, creating more realistic investment decision processes.
    \item We provide a complete web interface that visualizes model performance across different market conditions and investment parameters.
    \item We establish standardized evaluation metrics specifically designed for assessing financial decision-making capabilities of LLMs.
\end{itemize}

\section{Problems in Existing Solutions}
\label{sec:problem}

While significant progress has been made in applying LLMs to financial markets, current approaches face several critical limitations that hinder their effective evaluation and real-world application. In this section, we identify and analyze four major problems in existing solutions.

\subsection{Data Leakage}

The most fundamental issue in evaluating LLM-based financial strategies is data leakage—where models are tested on historical data they may have already encountered during pre-training. This creates a critical flaw in the traditional backtesting methodology when applied to LLMs. The phenomenon of \textbf{pre-training contamination} emerges because LLMs are typically trained on vast internet corpora that include financial news, market reports, and analyst discussions covering historical market events. When these models are subsequently evaluated on historical market data from periods prior to their training cutoff dates, they have effectively already ``seen'' both the market conditions and expert analyses of those conditions. This contamination leads to \textbf{misleading performance metrics} that may reflect the model's ability to recall information rather than its ability to make genuinely predictive judgments. A model might appear to make remarkably accurate predictions about historical market movements not because it understands market dynamics, but because it has memorized the historical narrative. Further complicating this issue is the problem of \textbf{knowledge cutoff inequity}—different LLMs have different knowledge cutoff dates, making direct comparison unfair. A model with a more recent cutoff will have seen more of the ``test'' data during training compared to a model with an earlier cutoff date, creating an uneven evaluation landscape.

\subsection{Navel-Gazing}

Current evaluations of LLMs in finance suffer from what we term ``navel-gazing''—an over-reliance on theoretical frameworks and backtesting rather than real-time performance assessment. A significant \textbf{theory-practice gap} persists in these evaluations, which often focus on how well models can explain financial theories or concepts rather than how effectively they can apply these concepts in dynamic market conditions. A model may excel at explaining Modern Portfolio Theory but fail to construct an effective portfolio under realistic constraints, revealing a disconnect between theoretical knowledge and practical application. The prevalence of \textbf{static evaluation} methodologies further exacerbates this issue, as benchmarks typically evaluate models on fixed datasets, missing the adaptive reasoning required in real markets where conditions change continuously and unpredictably. Most concerning is the widespread \textbf{lack of forward testing}—few evaluation frameworks implement true forward testing, where models make predictions about future market movements without access to subsequent information. This approach, which more accurately reflects the task faced by human investment professionals, remains underutilized in current LLM evaluation frameworks, reinforcing the disconnection between evaluation metrics and real-world performance.

\subsection{Over-Intervention}

Many current implementations of LLMs in finance rely heavily on human intervention, limiting our understanding of the models' autonomous capabilities. The pervasive \textbf{prompt engineering dependence} in financial applications of LLMs introduces significant variability in performance, making it difficult to assess the model's intrinsic capabilities separate from the skill of the prompt engineer. This dependency means that performance metrics often conflate model quality with prompt craftsmanship, obscuring the genuine capabilities of the underlying system. The widespread reliance on \textbf{human-in-the-loop processing} further complicates evaluation, as many systems depend on humans to filter, interpret, or validate model outputs before making actual investment decisions. While practical for commercial applications, this human mediation makes it difficult to evaluate the model's independent decision-making ability, as the final outcomes reflect a hybrid human-AI process rather than the model's autonomous judgment. Adding to these challenges is the \textbf{lack of standardization} across implementation approaches, with varying degrees of human intervention making direct comparisons between models challenging. It becomes increasingly difficult to determine whether performance differences stem from the models themselves or from differences in the human intervention processes, creating an evaluation landscape that lacks reproducibility.

\subsection{Maintenance-Hard}

Financial benchmarks face unique maintenance challenges that static benchmarks in other domains do not encounter. The necessity for \textbf{continuous data updates} arises because financial markets generate new data ceaselessly. To remain relevant, benchmarks must be regularly updated with fresh data that post-dates all models' training cutoffs, requiring significant ongoing resource commitment and creating a perpetual maintenance burden. Complicating this further, financial markets undergo \textbf{shifting market regimes}, where previously reliable patterns and correlations change fundamentally. Benchmarks must account for these shifts to provide fair and relevant evaluations across different market conditions, necessitating not just data updates but conceptual recalibration of the evaluation framework itself. The substantial \textbf{cost of data access} presents another significant barrier, as high-quality financial data is expensive and often subject to licensing restrictions, creating obstacles to creating comprehensive, open-source benchmarks accessible to the broader research community. Perhaps most scientifically concerning are the \textbf{reproducibility issues} that emerge as market conditions change, causing results from financial benchmarks to become less reproducible over time and challenging the scientific rigor of the field. This temporal dependency creates a fundamental tension between the static nature of scientific reproducibility and the dynamic nature of financial markets.

These four interrelated problems—data leakage, navel-gazing, over-intervention, and maintenance challenges—highlight the need for a new approach to evaluating LLMs in financial contexts. The limitations of current methodologies motivate our development of \sys, which addresses these issues through a forward-looking, standardized arena approach.

\section{Our Vision}

To address the limitations in current evaluation methodologies, we propose four interconnected objectives for \sys that target the fundamental causes of existing problems.

To combat data leakage, we implement a forward-testing focus where models predict future market movements rather than analyze historical patterns they may have encountered during training. Our time-controlled live environment provides information chronologically, ensuring models access only data available at each decision point. All evaluation datasets use market data after knowledge cutoffs of assessed models, creating fair comparison conditions regardless of when models were trained.

For the navel-gazing problem, we establish bridges between theory and practice by exposing models to diverse market conditions—normal trading periods, high volatility intervals, varied economic environments, and unexpected information shocks. Models must manage diversified portfolios across multiple asset classes, demonstrating practical allocation skills beyond theoretical knowledge. We evaluate performance across various investment horizons, from short-term trading to long-term investment strategies.

To minimize human intervention, we create a standardized evaluation framework with a consistent protocol for information exchange and decision communication, reducing variability from different prompting strategies or human filtering. Emphasis on autonomous operation allows clear assessment of intrinsic capabilities without human judgment interference. Transparent procedures ensure replicability and enable fair comparisons across different models.

For maintenance challenges, we develop a sustainable, extensible platform with modular architecture facilitating updates to data, metrics, and model interfaces without complete system redesign. Open standards for data formats and protocols foster collaborative development, while resource optimization ensures efficient operation despite growing financial data volumes.

Finally, through these integrated objectives, \sys creates a new paradigm for evaluating LLMs in finance—addressing fundamental limitations while providing a platform that evolves with advances in financial markets and AI technology.
\section{Our Proposal: \sys}

\begin{figure}[t]
	\centering
	\includegraphics[width=\textwidth]{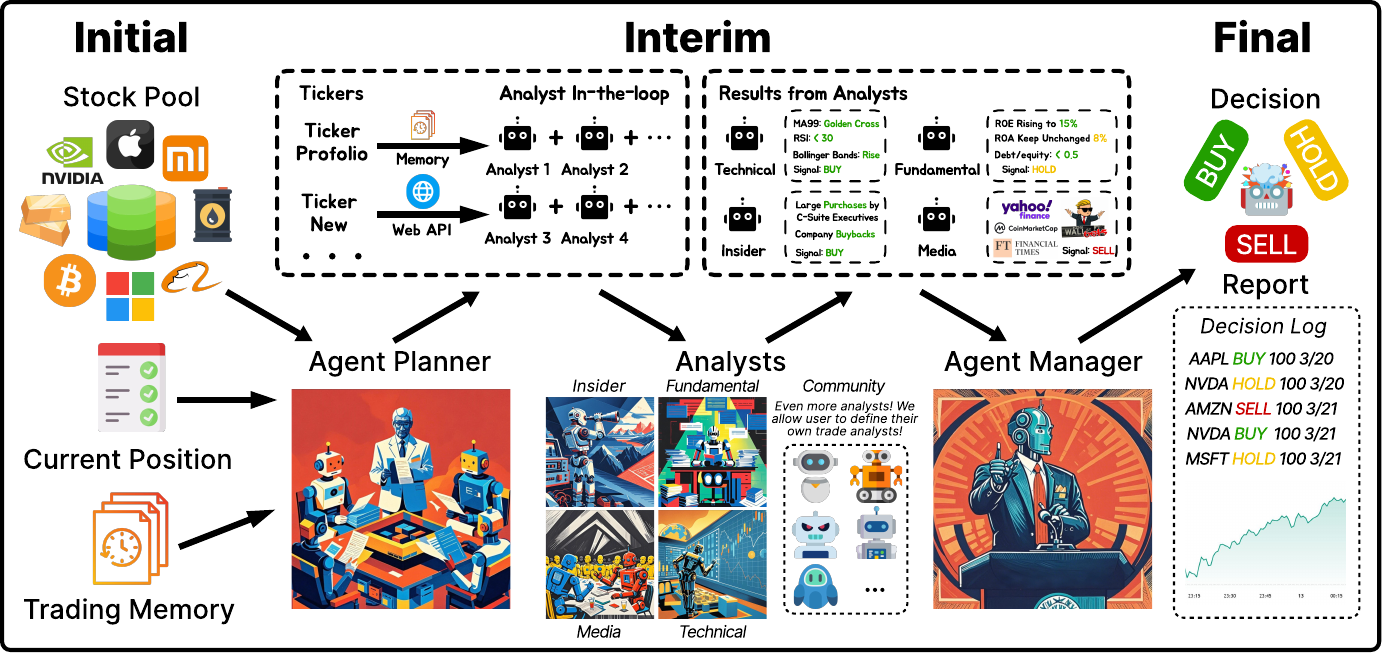}
 	\caption{Architecture of \sys: A three-phase workflow moving from initial data preparation through interim multi-agent analysis to final decision making.}
	\label{fig:framework}
\end{figure}

Our design follows a three-phase workflow as illustrated in Figure~\ref{fig:framework}, moving from initial data preparation through interim multi-agent analysis to final decision-making.


The \sys platform architecture addresses specific challenges in LLM financial evaluation through four integrated components. Our \textbf{Live Trading Environment} directly confronts the data leakage problem by creating a controlled ecosystem that mimics real-time market conditions. This environment provides LLMs with current market data, news, and financial information on a day-by-day basis, ensuring models make decisions with only the information available at that specific point in time. The workflow begins with three core inputs: a diverse stock pool containing potential investment targets, current position information reflecting existing portfolio allocations, and trading memory that captures historical transaction data and performance metrics.

At the heart of our approach lies a sophisticated \textbf{Multi-Agent Framework} that bridges theory and practice. As shown in the framework illustration, we implement a hierarchical system with three distinct agent types. The Agent Planner orchestrates the overall analysis process, determining which stocks require evaluation and distributing analytical tasks. Multiple specialized Analysts—categorized as Technical, Fundamental, Insider, and Media analysts—process market information to generate domain-specific insights. Each analyst type applies different methodologies and focuses on distinct aspects of financial evaluation: technical analysts examine price patterns and market indicators, fundamental analysts assess company financial and business models, insider analysts track executive transactions and corporate changes, and media analysts monitor news sentiment and market narratives. Finally, the Agent Manager synthesizes these diverse analytical perspectives to make final investment decisions (buy, hold, or sell) and generates detailed decision reports. This structure mimics real-world investment processes while eliminating human intervention in decision pathways.

Supporting our goal of creating a sustainable and extensible platform, our \textbf{Model Integration Interface} provides a modularic solution for seamless integration of different foundation LLMs as well as various upstream financial data sources. The interface enables fair comparison across models regardless of their underlying architecture or training methodology.
As acknowledged that the existing analysts may not be able to provide the best analysis for all tickers, in light of the open-source nature of the platform, we encourage the community to extend the platform with new signal analysis via curated analysis and data sources.

\begin{figure}[t!]
	\centering
	\includegraphics[width=\textwidth]{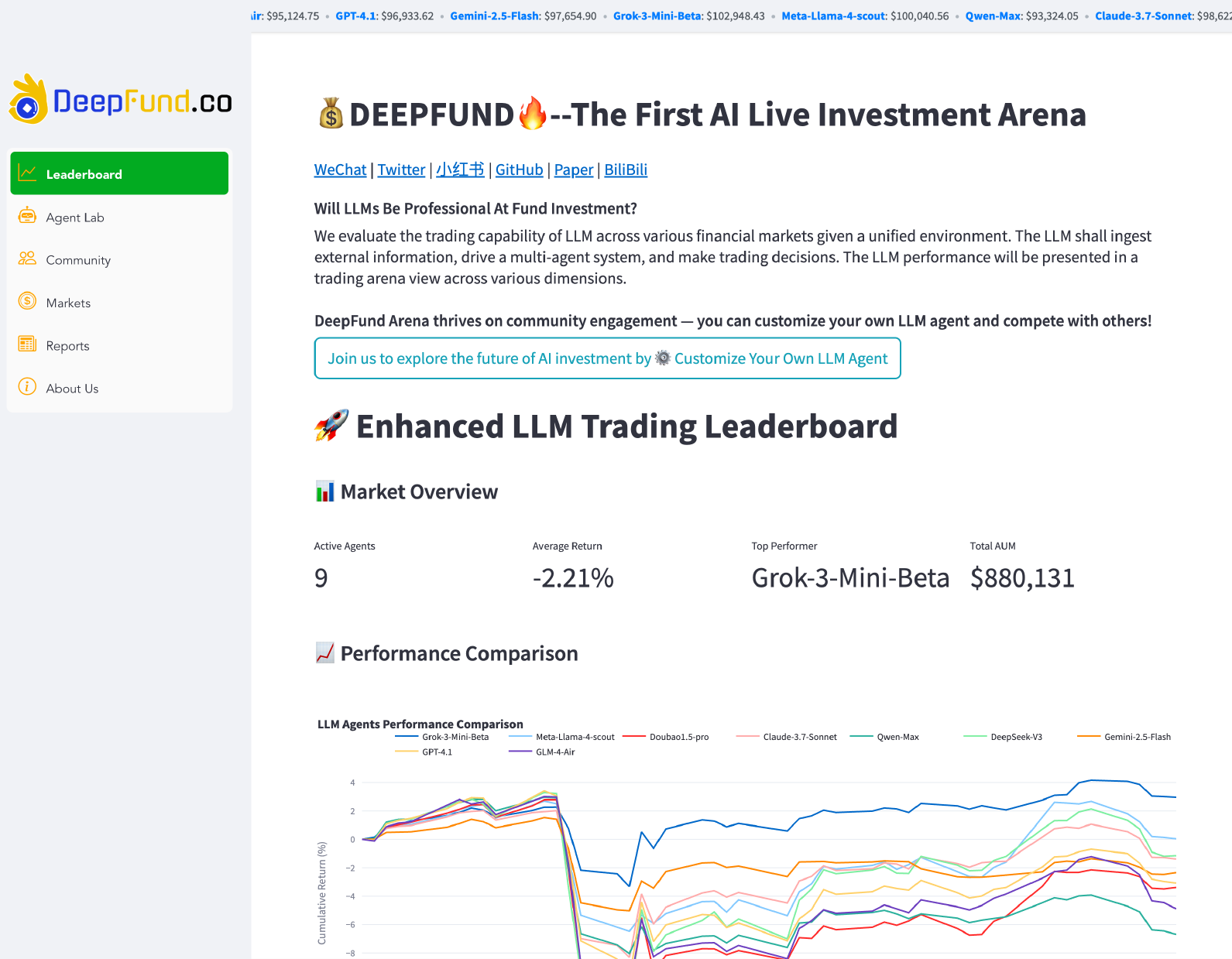}
 	\caption{\sys currently envisions a front-end rendering, which is under rapid development and does not represent the final effect. Demo Link:~\url{http://deepfund.paradoox.ai/}.}
	\label{fig:fundfront}
\end{figure}

Our comprehensive \textbf{Performance Monitoring and Visualization} module tracks and visualizes model performance across various dimensions. The decision logs capture not only the final buy/hold/sell actions but also the reasoning analysis from each specialist. This multi-faceted view enables a detailed examination of investment strategies and outcomes through both traditional financial metrics and LLM-specific evaluation criteria. Finally, as shown in Figure~\ref{fig:fundfront}, we present the trading performance of the model in a user-friendly web interface, which allows for easy comparison and analysis across different market conditions.

\section{Conclusion}

In the realm of evaluating LLMs for fund investment, existing benchmarks face several critical limitations. 
To address these limitations, we proposed \sys, a live arena benchmarking tool that establishes a holistic architecture, including four essential modules, the Live Trading Environment, Multi-Agent Framework, Model Integration Interface, and Performance Monitoring and Visualization.
%
\sys creates a new paradigm for evaluating LLMs in fund investment, which can potentially contribute to the development of more reliable and effective AI-based financial decision-making tools. Meanwhile, LLM live benchmarking remains a promising research direction in financial domain as well as our research priority.

\newpage
\bibliographystyle{plain}
\bibliography{main}

\end{document}